\documentclass[aps,prx,notitlepage,superscriptaddress,longbibliography,twocolumn,nofootinbib,floatfix]{revtex4-2} 

\usepackage[utf8]{inputenc}
\usepackage{graphicx}
\usepackage{hyperref}
\usepackage[dvipsnames]{xcolor}

\usepackage{pgfplots}

\usepackage{lipsum}

\usepackage{amsmath,amsthm,amssymb,mathtools}
\usepackage{multirow}
\usepackage{braket}
\usepackage{bbm,bm}
\usepackage[bb=boondox]{mathalfa}
\usepackage{dsfont}

\DeclareFontFamily{U}{matha}{\hyphenchar\font45}
\DeclareFontShape{U}{matha}{m}{n}{
      <5> <6> <7> <8> <9> <10> gen * matha
      <10.95> matha10 <12> <14.4> <17.28> <20.74> <24.88> matha12
      }{}
\DeclareSymbolFont{matha}{U}{matha}{m}{n}
\DeclareMathSymbol{\muparrow}{3}{matha}{"D2}
\DeclareMathSymbol{\mdownarrow}{3}{matha}{"D3}
\DeclareMathSymbol{\mupdownarrow}{3}{matha}{"D9}

\DeclareFontFamily{U}{mathb}{\hyphenchar\font45}
\DeclareFontShape{U}{mathb}{m}{n}{
      <5> <6> <7> <8> <9> <10> gen * mathb
      <10.95> mathb10 <12> <14.4> <17.28> <20.74> <24.88> mathb12
      }{}
\DeclareSymbolFont{mathb}{U}{mathb}{m}{n}
\DeclareMathSymbol{\mdownuparrows}{3}{mathb}{"D7}

\usepackage{makecell}

\usepackage{float}

\definecolor{dgreen}{rgb}{0,0.666,0}
\definecolor{dorange}{rgb}{0.666,.333,0}

\newcommand{\ie}{{\it i.e.},\ }

\newcommand{\Tr}{\operatorname{Tr}}

\makeatletter
\pgfmathdeclarefunction{erf}{1}{%
  \begingroup
    \pgfmathparse{#1 > 0 ? 1 : -1}%
    \edef\sign{\pgfmathresult}%
    \pgfmathparse{abs(#1)}%
    \edef\x{\pgfmathresult}%
    \pgfmathparse{1/(1+0.3275911*\x)}%
    \edef\t{\pgfmathresult}%
    \pgfmathparse{%
      1 - (((((1.061405429*\t -1.453152027)*\t) + 1.421413741)*\t 
      -0.284496736)*\t + 0.254829592)*\t*exp(-(\x*\x))}%
    \edef\y{\pgfmathresult}%
    \pgfmathparse{(\sign)*\y}%
    \pgfmath@smuggleone\pgfmathresult%
  \endgroup
}
\makeatother

\begin{document}
\title{Direct observation of dynamical quasi-condensation on a quantum computer
}

\author{Philipp Frey}
\affiliation{School of Physics, The University of Melbourne, Parkville, VIC 3010, Australia}

\author{Stephan Rachel}
\affiliation{School of Physics, The University of Melbourne, Parkville, VIC 3010, Australia}

\begin{abstract}
Hard-core bosons (HCB) in one dimension are predicted to show surprisingly interesting dynamics after a quantum quench. Far from equilibrium, quasi-condensation at finite momenta has been observed in numerical studies, while the equilibrium state at late times is expected to violate conventional thermodynamics. The integrability of the model supposedly constraints the momentum distribution to approach a generalized Gibbs ensemble. The experimental observation of these phenomena has proven non-trivial, as optical lattice platforms do not directly access the momentum distribution. NISQ devices overcome this limitation. We use circuit compression in order to simulate dynamics to arbitrarily long times with negligible Trotter-error on IBMQ and directly observe quasi-condensation. Coherence is maintained across all time scales as indicated by the lowest natural orbitals. The equilibrium distribution at late times is seemingly well described by a Gibbs ensemble, indicating that small but finite systematic errors perturb the hard-core boson model away from integrability. We demonstrate that quantum simulation provides observational access to HCB physics.
\end{abstract}

\maketitle


\clearpage


\section{Introduction}

\begin{figure*}[]
    \centering
    \includegraphics[width=1.05\linewidth]{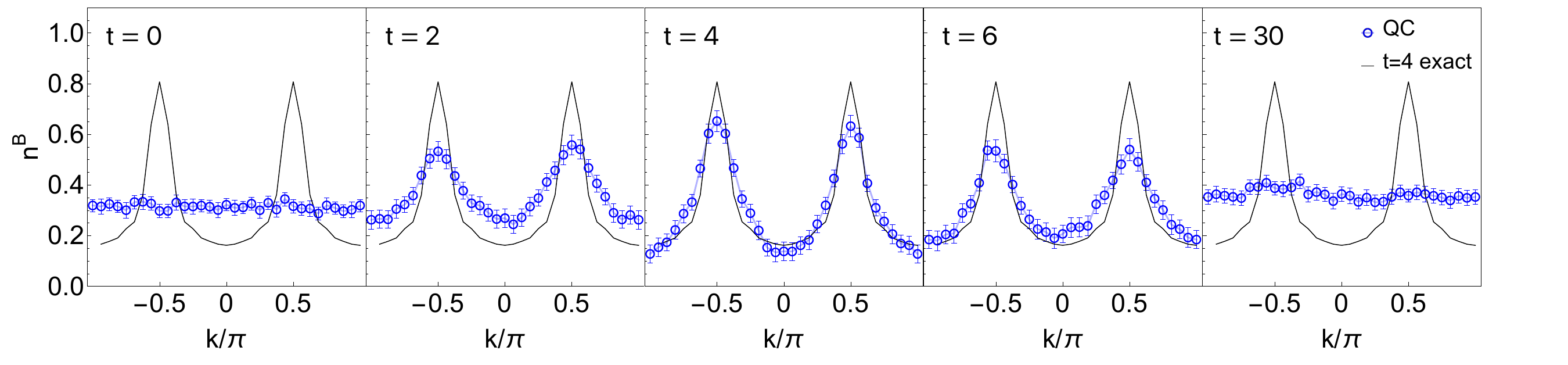}
    \caption{\textbf{Boson momentum distribution at different times after quench from Fock-state.} The solid black curve shows the exact numerical result for $t=4$. Two peaks form dynamically at momenta $k = \pm \pi/2$, where the dispersion vanishes, and decay at late times, demonstrating quasi-condensation.}
    \label{fig:boson-dist}
\end{figure*}

Hard-core bosons (HCB) in one dimension exhibit interesting dynamics, despite the fact that they can be mapped onto non-interacting spinless fermions with an exact analytic solution. This can be understood as a consequence of the non-locality of the mapping between these two types of particles. While certain quantities such as energy, local density and density-density correlations are identical in both descriptions, the single-particle correlations are typically quite different, which is represented in distinct momentum-distributions as well as the so-called natural orbitals (NO). For example, while the fermionic momentum distribution is conserved over time due to the non-interacting nature of the fermion description, the bosonic momentum distribution is time-dependent. Among the interesting phenomena expected in HCB dynamics are the dynamical formation of quasi-condensates\,\cite{PhysRevLett.93.230404} and the generalized thermalization at late times\,\cite{Vidmar_2016}.
The latter is a consequence of the integrability of the model and predicts equilibrium expectation values that may greatly differ from a conventional Gibbs ensemble.
Both phenomena can be studied numerically using an efficient algorithm for Gaussian states. Experimental realizations make use of optical lattices\,\cite{PhysRevLett.87.160405,PhysRevLett.91.250402,PhysRevLett.92.130403,PhysRevLett.92.190401,Paredes2004,doi:10.1126/science.1100700}, where the readout of arbitrary two-site correlation functions is a non-trivial task \cite{PhysRevLett.121.260401}. Instead, expansion velocities after a quench are obtained from the time-dependent real-space density\,\cite{PhysRevLett.110.205301}, or time-of-flight measurements are used to infer the momentum distribution. The experiments require an averaging over inequivalent 1D chains and suffer from imperfect initialization, as well as limitations due to the time-of-flight imaging technique that yields a convolution of momentum- and real-space density distributions\,\cite{PhysRevLett.115.175301}. A direct measurement of time-dependent momentum-space distributions across arbitrary time scales therefore has remained an open problem until now.

HCBs can also be mapped onto the spin-1/2 XY model\,\cite{LIEB1961407}, which has a natural implementation on current noisy intermediate-scale quantum (NISQ) devices. In order to perform the time evolution, the XY Hamiltonian has to be trotterized, i.e., an approximate discretization of the exact unitary can be implemented in terms of local single-qubit and two-qubit gates. The challenge of trotterized time evolution lies in the fact that a better approximation requires increased circuit-depth. Furthermore, longer time intervals require larger circuit-depth for a given accuracy of the approximation. The finite error rates on non-error corrected quantum computers therefore usually limit both the accuracy and the time-scales available for such simulations. However, it has been shown for the XY model and related ones that the Trotter circuit can be compressed algebraically to a finite depth that is linear in system size and independent of total time and Trotter precision\,\cite{doi:10.1137/21M1439298,BassmanOftelie2022,PhysRevA.105.032420,f3c}. This is due to a Yang-Baxter-type identity that allows for a reordering of any three matchgates that act pairwise and by turns on a set of three adjacent qubits. Compressing the circuit requires a classical overhead but allows for the elimination of Trotter errors and gives access to late-time dynamics for HCBs.
The effectiveness of this type of circuit compression was demonstrated for five qubits on IBMQ\,\cite{PhysRevA.105.032420}. In this work, we go beyond the proof-of-principle and apply the technique to a system of 32 qubits. In combination with error mitigation techniques such as zero noise extrapolation, this enables us to observe the dynamics of HCBs on arbitrary time-scales.
One of the advantages of NISQ devices over other experimental platforms is the single-cite resolution and  straightforward measurement of arbitrary correlation functions, which allows for direct access to the momentum distribution. Furthermore, the initial state can in principle be prepared exactly and consistently with vanishing entropy as opposed to the finite residual entropy found in optical lattice experiments\,\cite{Trotzky2010}. 

In Sec.~\ref{sec:theory} we introduce the HCB model and explain how time evolution and measurement of the relevant observables is achieved on a digital quantum computer. We present numerical and quantum simulation results on dynamical quasi-condensation in Sec.~\ref{sec: dynamical quasi-condensation} and on generalized thermalization in Sec.~\ref{sec:thermalization}. We end with a discussion in Sec.~\ref{sec:discussion}.

\section{Hard-Core bosons} \label{sec:theory}

Hard-core bosons on a one-dimensional lattice with $L$ sites are modeled by a tight-binding Hamiltonian with hopping amplitude $w$,

\begin{equation}\label{eq:HCBham}
    H=-w \sum_{j=1}^{L}\left(b_j^{\dagger} b^{\vphantom{\dagger}}_{j+1}+\text{H.c.}\right) .
\end{equation}
For simplicity we will assume periodic boundary conditions for the rest of this section, although open boundaries can be treated numerically in a completely analogous fashion. The operators $b_i^{\dagger}$ and $b_i$ obey bosonic commutation relations, with the exception of the anticommuation relations for two operators on the same site that enforce the hard-core constraint of no more than one boson on any lattice site,

\begin{equation}\label{eq:Comm}
\begin{aligned}
    \left[\hat{b}_i, \hat{b}_j^{\dagger}\right] = \left[\hat{b}_i, \hat{b}_j\right]=\left[\hat{b}_i^{\dagger}, \hat{b}_j^{\dagger}\right] &=0 \quad \mathrm{\forall \, i \neq j} \\
    \left\{\hat{b}_i, \hat{b}_i^{\dagger}\right\}=1, \quad \left(\hat{b}_i\right)^2=\left(\hat{b}_i^{\dagger}\right)^2 &=0  \; ,
\end{aligned}    
\end{equation}
where $[\cdot,\cdot]$ denotes a commutator and $\{\cdot,\cdot\}$ denotes an anti-commutator.
The Hamiltonian \eqref{eq:HCBham} represents a Hubbard model in the limit of infinite on-site repulsion and we will henceforth set the hopping amplitude $w=1$.
Ladder operators $S^+_i$ and $S^-_i$ describing spin-flips within a spin-1/2 chain obey the same (anti-) commutation relations as \eqref{eq:Comm}, i.e., spin-flips can be thought of as bosonic {\it particles}. Using $S^{\pm}_i \coloneq (X_i \mp i Y_i)/2$, the Hamiltonian \eqref{eq:HCBham} may be rewritten as 

\begin{equation}\label{XY}
   H= - \frac{1}{2} \sum_i\left(X_i X_{i+1} + Y_i Y_{i+1} \right) \;,
\end{equation} 
where $(X,Y,Z) = (\sigma^x,\sigma^y,\sigma^z)$. The connection to free fermions is established via the Jordan-Wigner transformation,

\begin{equation}
    b_i^{\dagger}=f_i^{\dagger} \; \prod_{\beta=1}^{i-1} e^{-i \pi f_\beta^{\dagger} f_\beta}, \quad b_i=\prod_{\beta=1}^{i-1} e^{i \pi f_\beta^{\dagger} f_\beta} \; f_i .
\end{equation}
The operators $f_i^{\dagger}$ and $f_i$ obey conventional fermionic anti-commutation relations,

\begin{equation}\label{eq:antiComm}
\begin{aligned}
     \left\{f_i, f_j\right\}= \left\{f_i^{\dagger}, f_j^{\dagger} \right\} &=0\ , \\
     \left\{\hat{f^\dagger}_i, f_j\right\} &= \delta_{ij} \mathds{1} ,
\end{aligned}    
\end{equation}
and the Hamiltonian \eqref{eq:HCBham} takes the form of a free fermion model,

\begin{equation}\label{Fham}
    H= - \sum_i\left(f_i^{\dagger} f^{\vphantom{\dagger}}_{i+1}+\text{H.c.}\right) .
\end{equation}
Unlike for the bosonic operators, the fermionic Hamiltonian \eqref{Fham} can be diagonalized by Fourier transformation of $f_j = 1/\sqrt{L} \sum_k \exp(i j k 2\pi/L ) \tilde{f}_k$:
\begin{equation}
    H =- 2 \sum_k \cos \left(k \frac{2\pi}{L} \right) \, \tilde{f}_k^{\dagger} \tilde{f}^{\vphantom{\dagger}}_k .
\end{equation}

This shows that the fermion momentum distribution $\tilde{n}^F_k = \langle  \tilde{f}_k^{\dagger} \tilde{f}^{\vphantom{\dagger}}_k \rangle $ is conserved and therefore represents an extensive set of conserved charges, making HCBs in one dimension an integrable model. The corresponding boson momentum distribution $\tilde{n}^B_k = \langle  \tilde{b}_k^{\dagger} \tilde{b}^{\vphantom{\dagger}}_k \rangle $ is, however, not conserved in general and distinct from the fermion distribution. We may express $\tilde{n}^B_k$ in terms of real-space correlations $\tilde{n}^B_k (t) = \sum_{l,m} e^{-i k(l-m) 2\pi/L} \langle b^\dagger_l b^{\vphantom{\dagger}}_m \rangle$ and use the fact the local densities are the same in both representations, i.e. $n^B_l = \langle b^\dagger_l b^{\vphantom{\dagger}}_l \rangle = \langle f^\dagger_l f^{\vphantom{\dagger}}_l \rangle =n^\mathrm{F}_l$, to see that the off-diagonal correlations are responsible for the difference between the two momentum distributions. In particular, off-diagonal long range order is associated with (quasi-) condensation in bosonic systems. In order to access the one-particle density matrix $\rho_{jl} = \langle b^\dagger_j b^{\vphantom{\dagger}}_l \rangle$ on a quantum computer, we use the mapping onto spins, i.e., qubits,

\begin{equation}
    \rho_{jl} =  \frac{1}{4} \left[ \langle X_j X_l \rangle +  \langle Y_j Y_l \rangle + i \langle X_j Y_l \rangle - i  \langle Y_j X_l \rangle \right] .
\end{equation}
The measurement of all pairwise $XX$, $YY$, and $XY$ correlators only requires $\mathcal{O}(\log L)$ distinct circuits. Our approach to observing HCB dynamics on a quantum computer will be to quench the system by preparing some non-equilibrium state and applying a compressed time evolution circuit, followed by the measurement of all expectation values that make up $\rho_{jl}$.

\begin{figure}[t!]
    \centering
    \includegraphics[width=1.0\linewidth]{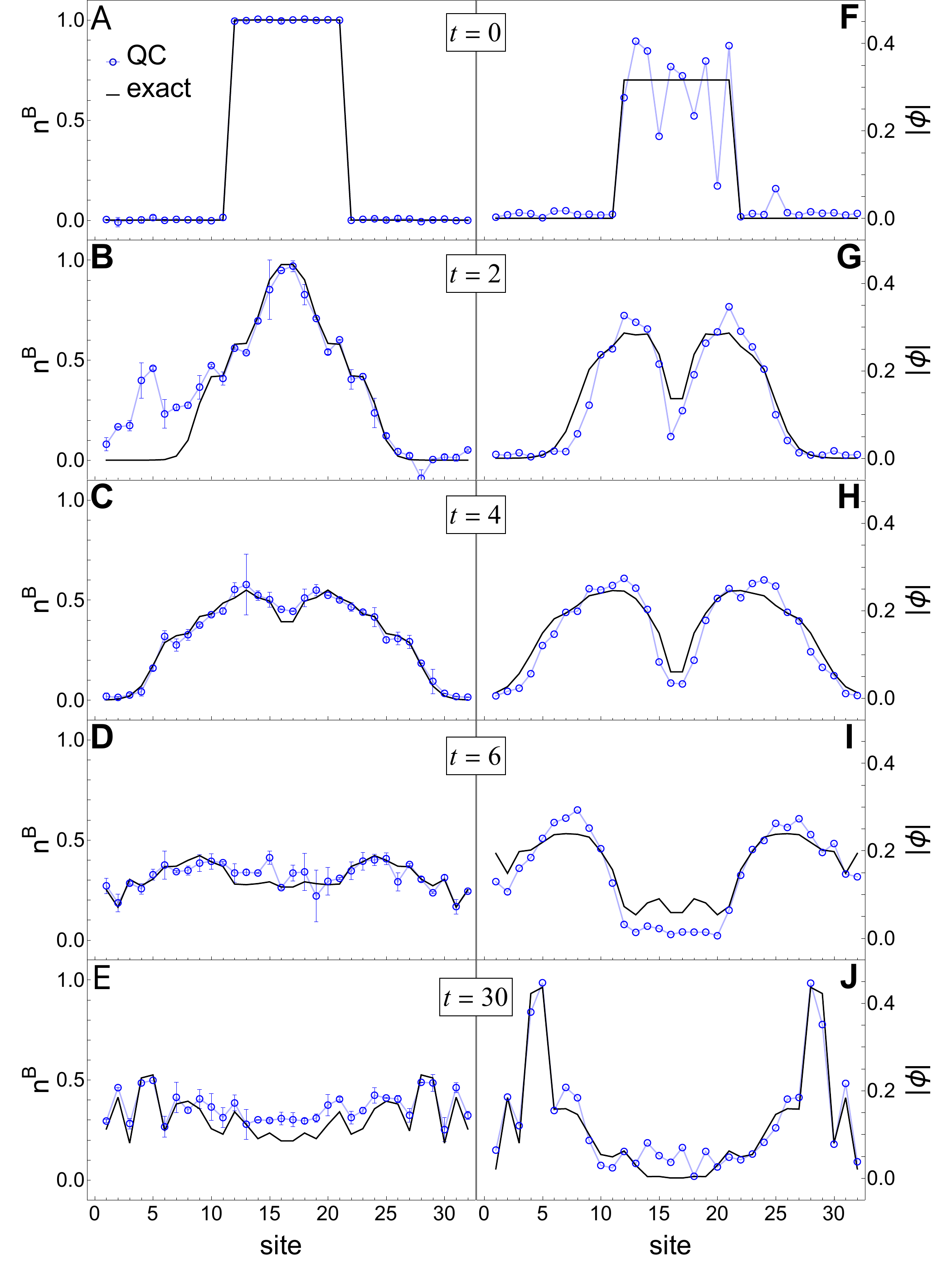}
    \caption{\textbf{Real-space density and lowest natural orbitals after a quantum quench.} The initial Fock state of $N=10$ bosons on $N$ lattice sites expands freely under $H$ into a box of $L=32$ sites. \textbf{(A-E)} real-space density at different times after the quench. At late times ($t=30$) the density is mostly homogeneous. \textbf{(F-J)} Lowest natural orbitals. The LNOs show two distinct lobes developing between $t=0$ and $t=6$, corresponding to the time-scale on which quasi-condensation is observed (see Fig\,\ref{fig:boson-dist}). Coherence is well maintained to late times ($t=30$), thanks to circuit compression.}
    \label{fig:QuasiLNO}
\end{figure}

\section{Dynamical quasi-condensation}\label{sec: dynamical quasi-condensation}

While the absence of Bose-Einstein condensation in one dimension precludes long range order in the one-particle density matrix, quasi-long range order in the sense of an algebraic decay with distance has been predicted. This corresponds to the occupation of few natural orbitals (NO), \ie single-boson wavefunctions, by $\mathcal{O}(\sqrt N)$ bosons. Quasi-condensation can be observed for a particularly simple initial state, namely a Fock-state of $N$ bosons in a box of size $Na$, with $a$ representing the lattice constant, placed in the center of the lattice. This state may be prepared by applying $X$-gates on the corresponding qubits after initialization in the $\ket{0}$ state. Applying a compressed Trotter circuit to this initial state implements free expansion within a box of size $L$ up to some time $t$, measured in units of $\hbar/w$. The initial fermion and boson momentum distributions for a Fock-state are identical and $k$-independent due to vanishing off-diagonal correlations. After a finite time the boson distribution~$\tilde{n}^B_k$ develops two distinct peaks at momenta $k =\pm \pi/2$, as shown in Fig.\,\ref{fig:boson-dist}. In addition, we compare the lowest natural orbitals (LNO) obtained from the quantum simulation to exact numerical results in Fig.\,\ref{fig:QuasiLNO}. The LNO is the highest occupied one-particle mode defined as an eigenvector of the one-particle density matrix,

\begin{equation}
    \sum_{j=1}^L \rho_{i j}(t) \phi_j^n(t)=\lambda_n(t) \phi_i^n(t)  .
\end{equation}
Numerically, we find that there are two degenerate largest eigenvalues corresponding to even and odd superpositions of a left-moving and a right-moving mode, respectively\,\cite{PhysRevLett.93.230404}. Due to noise, the corresponding two largest eigenvalues obtained from our quantum simulation are non-degenerate. We consider superpositions of the two eigenfunctions and minimize the norm of the difference between the measured orbital and the exact one obtained numerically. 
Both the real-space density and the LNOs as measured on \textit{ibm\_fez} follow the exact results very closely. The apparent discrepancy between the LNOs at $t=0$ may be attributed to the exact degeneracy of the eigenvalues in the case of a Fock-state. Here, the eigenvectors are determined by arbitrarily weak noise.
Clearly visible are two lobes in the LNOs at intermediate times, corresponding to a right-moving and a left-moving mode, respectively. We also show the real-space density as a function of time. At late times ($t=30$), the overall density flattens out. However, the LNOs maintain coherence as they oscillate back and forth and the data obtained from the QC tracks the dynamics across all time-scales. This suggests that long time scales are indeed accessible when using to circuit compression and that we should be able to probe the equilibrium distribution of the system instead of a thermal state brought about by decoherence.

\section{Generalized thermalization} \label{sec:thermalization}

Closed ergodic systems are expected to dynamically thermalize at sufficiently late times. This notion can be made precise by considering the time evolution of the reduced density operator $\rho_S$ for a sufficiently small local subsystem $\mathcal{S}$,

\begin{equation}
    \lim _{\substack{t \rightarrow \pm \infty \\ \bar{\mathcal{S}} \rightarrow \rightarrow \infty}} \rho_\mathcal{S}(t)=\rho_{\mathcal{S}}^{(\mathrm{eq})}(T, \mu, \ldots)=\operatorname{Tr}_{\bar{\mathcal{S}}}\left[e^{-\beta(H-\mu N+\ldots)} / Z\right]  ,
\end{equation}

\noindent where $\bar{\mathcal{S}}$ represents the complement of $\mathcal{S}$ and $Z$ is the partition function. The expectation values of local observables at late times are identical to those obtained from an equilibrium Gibbs ensemble (GE). The ensemble is constrained by the initial expectation values of all conserved charges, such as energy and particle number, by means of a corresponding Lagrange multiplier, such as temperature and chemical potential. In contrast to generic quantum chaotic systems, integrable models have an extensive set of conserved charges. Hence, the late time equilibrium is much more constrained and described by a generalized Gibbs ensemble (GGE) \cite{Vidmar_2016}. For the HCB model the set of integrals of motion is given by the fermionic momentum modes $\tilde{n}^F_k$. The GGE is therefore given by

\begin{equation}\label{eq:GGE}
    \hat{\rho}_{\mathrm{GGE}}=Z_{\mathrm{GGE}}^{-1} \exp \left[-\sum_k \lambda_k \tilde{n}^\mathrm{F}_k \right]   ,
\end{equation}
where $Z_{\mathrm{GGE}} = \Tr[\hat{\rho}_{\mathrm{GGE}}]$. The Lagrange multipliers $\lambda_k$ are fixed by the initial conditions  \mbox{$\langle \tilde{n}^\mathrm{F}_k \rangle_0 =  \Tr[ \tilde{n}^\mathrm{F}_k \hat{\rho}_{\mathrm{GGE}}]$}, where $\langle \cdot \rangle_0$ denotes the expectation value in the initial state. Straightforward calculation gives $\lambda_k=\ln \left(\frac{1-\langle\tilde{n}^\mathrm{F}_k\rangle_0}{\langle\tilde{n}^\mathrm{F}_k\rangle_0}\right)$ and $Z_{\mathrm{GGE}}^{-1} = \prod_k (1-\langle\tilde{n}^\mathrm{F}_k\rangle_0)$. In contrast, the conventional Gibbs ensemble for this model would only take into account the conservation of energy, $ H$, and total particle number, $N=\sum_k \tilde{n}^\mathrm{F}_k$, \ie

\begin{equation}\label{eq:GE}
    \hat{\rho}_{\mathrm{GE}}=Z_{\mathrm{GE}}^{-1} \exp \left[-\beta\left( H - \mu N \right)\right]   \; .
\end{equation}
Here, $\beta = 1/(k_B T)$ is the inverse temperature, $\mu$ the chemical potential and $Z_{\mathrm{GE}}=\Tr[\hat{\rho}_{\mathrm{GE}}] = \prod_k \left(1+e^{-\beta(\epsilon_k-\mu})\right)$, with $\epsilon_k$ denoting the single-particle spectrum of $H$.

\begin{figure}[t!]
    \centering
    \includegraphics[width=1.0\linewidth]{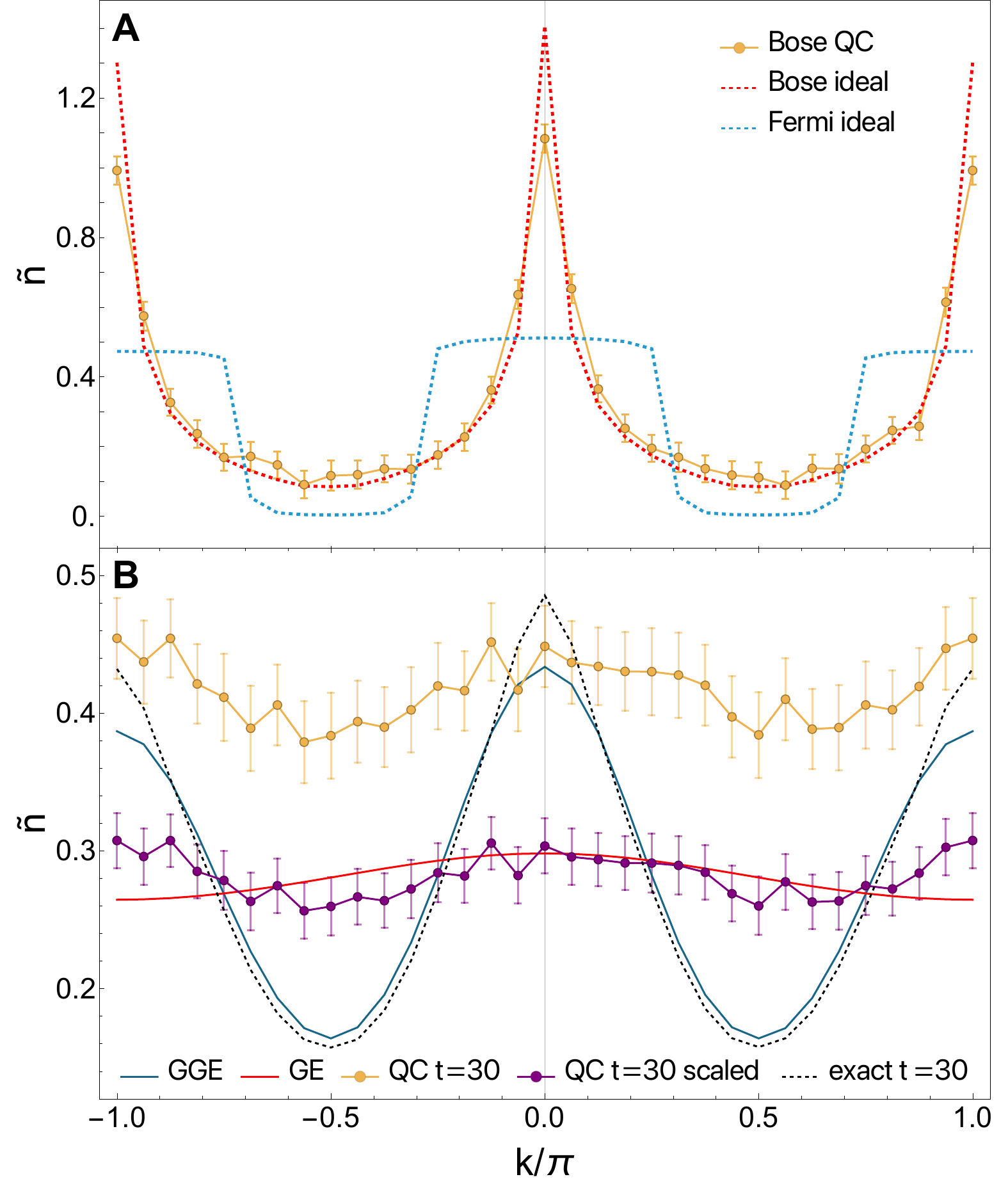}
    \caption{\textbf{Late-time equilibrium distribution.} \textbf{(A)} Bosonic (dashed red) and fermionic (dashed blue) momentum distributions of the ideal initial state as computed with exact numerics, along with the measured bosonic distribution (orange) after initialization on \textit{ibm\_fez}. \textbf{(B)}
    Comparison of ideal (dashed black) and measured (orange) late-time ($t=30$) bosonic momentum distributions with the predicted equilibrium based on the conventional Gibbs ensemble (red) and generalized Gibbs ensemble (dark blue). The proabilistic error amplification overestimates depolarization and a rescaling of the expecation values to the correct (conserved) particle number produces the purple curve.} 
    \label{fig:GGE}
\end{figure}

\subsection{State preparation}

A Fock-state such as the one used to observe dynamical quasi-condensation in Sec.\,\ref{sec: dynamical quasi-condensation} would have a flat fermionic momentum distribution and therefore a flat GGE, as well as a flat GE since it represents an infinite temperature state. We require non-vanishing off-diagonal correlations $\langle f^\dagger_j f^{\vphantom{\dagger}}_{l}\rangle_0 \neq 0$ in our initial state. The addition of a large superlattice potential with period $p$,  

\begin{equation}
     H' = \sum_{j=1}^L -\left(b_j^{\dagger} b^{\vphantom{\dagger}}_{j+1}+\text{H.c.}\right) + V \cos \left(j \frac{2\pi}{L} p \right)  b_j^{\dagger} b^{\vphantom{\dagger}}_{j} \; ,
\end{equation}
where $V\gg 1$ and $p$ divides $L$, gives rise to a low-energy sector whose fermionic single particle eigenmodes are approximately given by

\begin{equation}\label{eq:eigenmodes}
    c^\dagger_q \approx \sqrt{\frac{p}{L}} \sum_{n=1}^{L/p} e^{i m q \frac{2\pi}{L/p}} f^\dagger_{np} \quad , q=0,\dots,L/p-1 \;.
\end{equation}
The dispersion is given by $\epsilon_q = -(V+2\cos(q 2\pi/L p))$. One may easily verify that the contribution of each $c^\dagger_q$ to the fermion momentum distribution $\tilde{n}^F_k = \langle  \tilde{f}_k^{\dagger} \tilde{f}_k \rangle $ is $1/p \sum_k \delta_{(q-k)\mod(L/p),0}$, \ie $p$ equally spaced delta-peaks of amplitude $1/p$. The corresponding contribution to the HCB momentum distribution is more complicated due to the non-trivial correlations when more than one mode is filled. 
The many-body ground state $\ket{\Psi_0}$ for $N<L/p$ has non-trivial off-diagonal correlations due to the delocalized nature of these eigenmodes. In Fig.~\ref{fig:GGE}A we show the case of $N=L/(2p)+1$, which seems to maximize the amplitude of peaks in the distribution. 
This Gaussian state may be expressed as a Slater determinant,

\begin{equation}
    \ket{\Psi_0}=\prod_{n=1}^{N} \sum_{i=1}^L P_{i n}f_i^{\dagger}\ket{0}  .
\end{equation}
While the matrix elements $P_{i n}$ may be taken to be the coefficients in Eq.\,\eqref{eq:eigenmodes}, they are not uniquely determined. We can make use of this ambiguity and choose a different set of modes that are constructed by a sequence of Givens rotations, starting from a state of $N$ occupied lattice sites, $\ket{\Psi'_0}= \prod_{i=1}^N f^\dagger_{j_i} \ket{0}$. Each transformation can be implemented as a local two-qubit gate, allowing us to prepare the ground state $\ket{\Psi_0}$ on the quantum computer.

\subsection{Late-time dynamics}

While we expect to see GGE at late times due to integrability in case of the exact HCB model, it has been shown that gate errors on NISQ devices can result in systematic contributions to the effective Hamiltonian that can break integrability \cite{doi:10.1126/sciadv.abm7652}. The fact that the measured distributions in Fig.\,\ref{fig:QuasiLNO}F-J are in good agreement with numerics across all time-scales indicates that coherence is sufficiently maintained and significant deviation from GGE should not be attributed to equilibration with the environment. Fig.\,\ref{fig:GGE}A shows the ideal boson and fermion momentum distributions for the initial state, as well as the bosonic distribution obtained on the quantum computer, closely matching the ideal distribution. The same compressed time evolution circuit for $t=30$ is then applied to the prepared state and the required correlations are measured. The noise mitigation protocol includes probabilistic error ammplification (PEA), measurement error mitigation and gate twirling. Fig.\,\ref{fig:GGE}B shows the error mitigated data including standard deviation. We find that PEA appears to over-correct for depolarization and overestimates the total number of particles. We therefore also show rescaled data that corresponds to the correct filling. In comparison we show the GE and GGE predictions computed numerically from the exact initial state, as well as the boson momentum distribution of the exact time-evolved state. The latter is in good agreement with the GGE distribution, while the QC data follows the GE distribution. This indicates that systematic errors perturb the Hamiltonian sufficiently far from the integrable point as to change the qualitative dynamics from integrable to quantum chaotic. One can observe a mild modulation in the unscaled data, visible in Fig.\,\ref{fig:GGE}B (orange data points). This modulation seems to follow the GGE prediction. While the effect is rather weak, one might conjecture that it indicates an intermediate distribution.

\section{Discussion}\label{sec:discussion}

To summarize, we have successfully implemented quantum quenches of two different initial states for the integrable hard-core boson model in one dimension on a qauntum computer. Using algebraic circuit compression, we were able to observe coherent dynamics across arbitrary time scales as demonstrated by the measurement of the lowest natural orbitals and comparison to exact numerics. Using the ability to measure arbitrary correlation functions we directly obtained the time-dependent boson momentum distribution, which allowed us to observe dynamical quasi-condensation at short to intermediate times, as well as the late-time equilibrium distribution. The latter was theoretically and numerically predicted to follow a generalized Gibbs ensemble, as opposed to a conventional Gibbs ensemble, in the ideal case. We successfully initialized a Gaussian state for which the GGE and GE predictions significantly deviate from each other and measured the distribution of a time-evolved state. Our results suggest that integrable dynamics is highly sensitive to small systematic errors that shift the effective Hamiltonian slightly away from the exact HCB model and push the late-time dynamics towards conventional thermalization. However, the raw data shows a modulation that qualitatively follows the GGE prediction. This would suggest that current NISQ devices are on the verge of being able to distinguishing integrable thermodynamics from conventional thermodynamics, a task that could serve as a challenging benchmark for near-future quantum computing devices.

\section*{Acknowledgments}
The authors acknowledge valuable discussions with Michael Jones and Lloyd Hollenberg. This work was supported by the University of Melbourne through the establishment of an IBM Quantum Network Hub at the University. S.R. acknowledges support from the Australian Research Council through Grants No. DP200101118 and DP240100168.

\newpage

\clearpage
\newpage
\bibliography{references}

\clearpage
\onecolumngrid
\appendix

\section{Classical algorithm for HCB dynamics and (generalized) Gibbs ensembles}\label{App:HCBsimulation}

Here, we briefly discuss the classical algorithms that generate the exact numerical results in this work. We largely follow Rigol et al. \cite{doi:10.1142/S0217984905008876,PhysRevA.72.063607}. 

For a given quadratic Hamiltonian $H_F$ one may find the exact single-particle eigenstates $P_{i n}$ numerically, where the index $n$ labels the states in ascending order of energy and $i$ is the lattice index. The $N_F$-body ground state is then the Gaussian state 

\begin{equation}\label{eq:ground state}
    \ket{\Psi_F^I}=\prod_{n=1}^{N_F} \sum_{i=1}^L P_{i n}f_i^{\dagger}\ket{0}  .
\end{equation}

This state may be time-evolved by evolving the matrix $P_{i n}(t)$, e.g. via Fourier transform and multiplication by phase factors in the case of PBC,

\begin{equation}
    \ket{\Psi(t)}=e^{-i H_F t}\ket{\Psi^I}=\prod_{n=1}^{N_F} \sum_{i=1}^L P_{i n}(t) f_i^{\dagger}|0\rangle  .
\end{equation}
The HCB Green's function can now be written in terms of fermionic operators using the Jordan-Wigner transformation,

\begin{equation}
    G_{i j}(t)=\left\langle\Psi(t)\left|b^{\vphantom{\dagger}}_i b_j^{\dagger}\right| \Psi(t)\right\rangle = \left\langle\Psi(t)\left|\prod_{l=1}^{i-1} e^{i \pi f_l^{\dagger} f^{\vphantom{\dagger}}_l} \, f^{\vphantom{\dagger}}_i f_j^{\dagger} \, \prod_{m=1}^{j-1} e^{-i \pi f_m^{\dagger} f^{\vphantom{\dagger}}_m}\right| \Psi(t)\right\rangle .
\end{equation}
The action of $ \prod_{m=1}^{j-1} e^{-i \pi f_m^{\dagger} f_m}$ on $\ket{\Psi(t)}$ is to change the sign of all entries in $P_{n i}(t)$ with $n < j$. The action of $f_j^{\dagger}$ is then to add a column to the matrix $\mathbf{P}(t)$ with the only non-zero entry given by $P_{j N_F+1}=1$. Analogous changes are affected by $\prod_{l=1}^{i-1} e^{i \pi f_l^{\dagger} f_l} \, f_i$ on the left, such that the result is given by

\begin{equation}\label{eq:Green}
G_{i j}(t)  =\left\langle 0\left|\prod_{n=1}^{N_F+1} \sum_{i=1}^N P_{i n}^{\prime A}(t) f^{\vphantom{\dagger}}_i \prod_{m=1}^{N_F+1} \sum_{j=1}^N P_{j m}^{\prime B}(t) f_j^{\dagger}\right| 0\right\rangle 
 =\operatorname{det}\left[\left(\mathbf{P}^{\prime A} (t)\right)^{\dagger} \mathbf{P}^{\prime B} (t)\right] ,
\end{equation}
where $\mathbf{P}^{\prime A} (t)$ and $\mathbf{P}^{\prime B} (t)$ are the two modified matrices. The last equality follows from the properties of the fermionic creation and annihilation operators.
Finally we see that the one-particle density matrix is related to $G$ via

\begin{equation}
    \rho_{i j}(t)=\left\langle\Psi(t)\left|b_i^{\dagger} b^{\vphantom{\dagger}}_j\right| \Psi(t)\right\rangle=G_{j i}(t)+\delta_{i j}\left[1-2 G_{i i}(t)\right] ,
\end{equation}
because of the (anti-)commutation relations \eqref{eq:Comm} of the HCB operators.

The bosonic one-particle density matrix of the generalized Gaussian ensemble for a given initial state $\Psi$ may be computed using the identity 

\begin{equation}\label{eq:identity}
 \operatorname{Tr}\left[\exp \left(\sum_{i j} f_i^{\dagger} X^{\vphantom{\dagger}}_{i j} f^{\vphantom{\dagger}}_j\right) \exp \left(\sum_{k l} f_k^{\dagger} Y^{\vphantom{\dagger}}_{k l} f^{\vphantom{\dagger}}_l\right) \cdots \exp \left(\sum_{m n} f_m^{\dagger} Z^{\vphantom{\dagger}}_{m n} f^{\vphantom{\dagger}}_n\right)\right] 
=\operatorname{det}\left[\mathbf{I}+e^{\mathbf{X}} e^{\mathbf{Y}} \cdots e^{\mathbf{Z}}\right]  .
\end{equation}
Here, we will assume PBC for simplicity. The expectation values of interest are given by the expression
\begin{equation}\label{eq:1pGGE}
    \langle\rho_{ij}\rangle = \frac{1}{Z_\mathrm{GGE}} \Tr \left[ f_i^\dagger f^{\vphantom{\dagger}}_j \prod_{l=1}^{j-1} \exp \left(i \pi n^F_l \right) \exp \left[ -\sum_k \lambda_k \tilde{n}^F_k \right] \prod_{m=1}^{i-1} \exp \left(-i \pi n^F_m \right)  \right]  ,
\end{equation}
where the partition function is $Z_\mathrm{GGE} = \Tr \left[ \exp \left(-\sum_k \lambda_k \tilde{n}^F_k \right) \right] = \prod_k \left(1+e^{-\lambda_k} \right)$.
Using the identity (valid for $i \neq j$) $f_i^{\dagger} f_j=\exp \left(\sum_{m n} f_m^{\dagger} A_{m n} f_n\right)-1$ and \eqref{eq:identity}, we find that for $i \neq j$ this simplifies to 

\begin{equation}
    \langle\rho_{ij}\rangle = \frac{1}{Z_\mathrm{GGE}} \left\{ \det \left[ \mathbb{I} + (\mathbb{I} + A) O_1 e^{-X} O_2  \right] - \det \left[  \mathbb{I} +   O_1 e^{-X} O_2 \right] \right\} ,
\end{equation}
where $\mathbb{I}$ is the $L \times L$ identity matrix, $A$ has only one non-zero entry $A_{ij}=1$, $O_1$($O_2$) is diagonal with the first $j-1$ ($i-1$) entries equal to $-1$ and the rest equal to $+1$. Finally, the matrix $X$ is defined such that $\sum_k \lambda_k \tilde{n}^F_k = \sum_{m,n} f^\dagger_m X^{\vphantom{\dagger}}_{mn} f^{\vphantom{\dagger}}_n$, \ie $X_{mn}=1/L \sum_k \lambda_k e^{i k (m-n) 2 \pi/L}$.
For $i=j$, the Jordan-Wigner strings cancel in \eqref{eq:1pGGE} and we can compute the fermionic expectation value $\langle f^\dagger_i f^{\vphantom{\dagger}}_j \rangle$ by introducing chemical potentials into the partition function 

\begin{equation}
    Z_\mathrm{GGE}(\{\mu_m\})= \Tr \left[ \exp \left( -  \sum_{m,n} f^\dagger_m \left(X^{\vphantom{\dagger}}_{mn} - \mu^{\vphantom{\dagger}}_m \delta^{\vphantom{\dagger}}_{mn} \right) f^{\vphantom{\dagger}}_n  \right) \right] = \det \left( \mathbb{I}+\exp \{-(X-\operatorname{diag}(\{\mu_m\})\} \right) . 
\end{equation}
From here we can use Jacobi's formula to get 
\begin{equation}
    \langle f^\dagger_i f^{\vphantom{\dagger}}_i \rangle = \frac{1}{Z_\mathrm{GGE}} \partial_{\mu_i} Z_\mathrm{GGE} \bigg\rvert_{\mu = 0} = \sum_{l,m,n} \left[ \mathbb{I} + e^{-X} \right]^{-1}_{lm} \left[e^{-X}\right]_{mn} P^{(i)}_{nl} ,
\end{equation}
where $P^{(i)}_{nl} = \delta_{ni}\delta_{li}$. We finally find

\begin{equation}
     \langle\rho_{ii}\rangle = \left( \left[ \mathbb{I} + e^{-X} \right]^{-1} e^{-X} \right)_{ii} = \left( V^\dagger \left[ \mathbb{I} + e^{-\Lambda} \right]^{-1} e^{-\Lambda} V \right)_{ii} =\left( V^\dagger \operatorname{diag}(\{\langle \tilde{n}^F_k \rangle\}) V \right)_{ii} ,
\end{equation}
where $\Lambda = \operatorname{diag}(\{\lambda_k\})$ and $V X V^\dagger = \Lambda$, which is accomplished by $V_{km}=1/\sqrt{L}e^{-i k m 2 \pi/L}$.

The corresponding result for a Gaussian ensemble with the same intial state $\Psi$ may be computed by first finding energy $\bar{E} = \bra{\Psi} H \ket{\Psi}$ and particle number $\bar{N}=\bra{\Psi} N \ket{\Psi}$, then the appropriate inverse temperature $\beta$ and chemical potential $\mu$, such that $\beta^{-1} \partial_\mu \log Z_\mathrm{GE} = \bar{N}$ and $-\partial_\beta \log Z_\mathrm{GE} + \mu \bar{N}= \bar{E}$. The rest of the calculation is analogous to the GGE derivation above with the expression $\exp \left[ -\sum_k \lambda_k \tilde{n}^F_k \right]$ in \eqref{eq:1pGGE} being replaced by $\exp \left[ -\sum_k \beta ( \epsilon_k  - \mu) \tilde{n}^F_k \right]$, where $\epsilon_k$ denotes the single-particle spectrum of $H_F$. 

The case of open boundary conditions (OBC) is analogous to the calculations presented above, except that the matrix elements of $V$ are to be replaced by the eigenvectors of the non-periodic Hamiltonian and $k$ no longer labels momenta but is merely the index of eigenstates.

\section{Fermionic N-particle Gaussian state on a quantum computer}\label{App:Gaussian}

\begin{figure}
    \centering
    \includegraphics[width=0.5\linewidth]{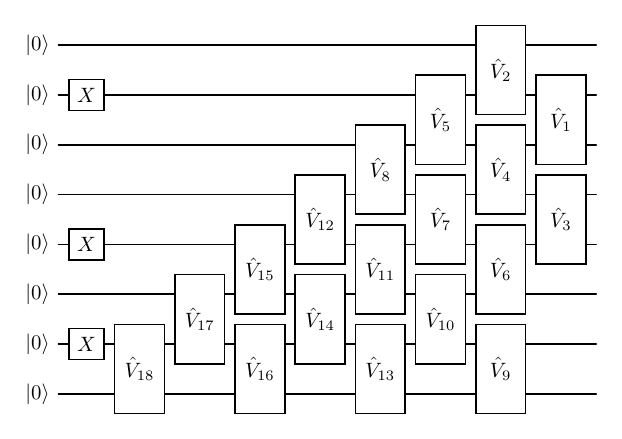}
    \caption{Illustration of the structure of a circuits comprised of $X$-gates and many-body Givens rotations that map the initial state onto an $N_F$-particle Gaussian state on $L$ sites. The indices indicate the order in which the corresponding rotations were applied to the eigenvectors in order to arrive at a diagonal correlation matrix. Here, $L=8$ and $N_F=3$.}
    \label{fig:InitStateGivens}
\end{figure}

\begin{figure}
    \centering
    \includegraphics[width=0.5\linewidth]{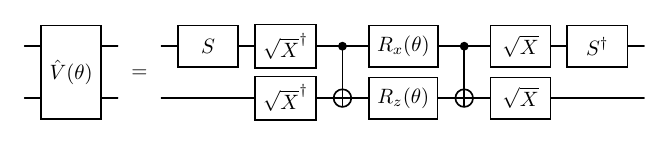}
    \caption{Decomposition of a Givens rotation acting on two qubits in terms of single-qubit and CNOT gates.}
    \label{fig:GivensDecomp}
\end{figure}

We outline the process of constructing the ground state of a fermionic quadratic Hamiltonian $H_F$ following an approach presented in \cite{PhysRevB.92.075132}. First we write the many-body Hamiltonian explicitly as $H_F= \sum_{m,n} f^\dagger_m h^{\vphantom{\dagger}}_{mn} f^{\vphantom{\dagger}}_n$. The many-body ground state $\ket{0}$ satisfies $f_n \ket{0} = 0$ for all $n$. We then diagonalize the matrix $h$ and write the $N_F$-particle ground state as in Eq.\,\eqref{eq:ground state}, the columns of $P$ being the eigenvectors of $h$. From here we compute the correlation matrix

\begin{equation}
    \Lambda_{i j}=\left\langle\Psi_F^I\left|f_i^{\dagger} f^{\vphantom{\dagger}}_j\right| \Psi_F^I\right\rangle=\delta_{i j} - G^F_{ji} ,
\end{equation}
where $G^F_{ji} = \left\langle\Psi_F^I\left|f^{\vphantom{\dagger}}_j f_i^{\dagger}  \right| \Psi_F^I\right\rangle$ is computed in the same manner as Eq.\,\eqref{eq:Green} but without the sign changes from the Jordan-Wigner strings. We now consider the partial correlation matrix $\Lambda_{i' j'}$ with $i',j'=\{1,\dots,n\}$ and iteratively increase $n$ starting from $n=1$ until one of its eigenvalues equals 0 or 1 within a specified precision. The corresponding eigenvector takes the form $(v_1, v_2, \dots, v_n)$ and is also an approximate eigenvector of the full matrix $\Lambda$ (when extended with $L-n$ entries of 0). We can map the eigenvector onto the fully localized  $(1, 0, \dots,0)$ by a sequence of Givens rotations

\begin{equation}
    V_i = \begin{bmatrix}
        \cos \theta_i & -\sin \theta_i \\
        \sin \theta_i & \hphantom{-}\cos \theta_i
    \end{bmatrix}       \quad,
\end{equation}
acting in the plane $(i,i+1)$. The correlation matrix transforms as 

\begin{equation}
    \Lambda' = V_{1} \cdots V_{n-1} \Lambda V^\dagger_{n-1} \cdots V^\dagger_{1} \;.
\end{equation}
We then take the partial $\Lambda'_{i' j'}$ with $i',j'=\{2,\dots,n\}$ and follow the same procedure starting with $n=2$. Eventually we arrive at the diagonal matrix

\begin{equation}
    \tilde{\Lambda} = V \Lambda V^\dagger = \operatorname{diag}(0,\dots,0,1,0,\dots,0,1,0,\dots) \;,
\end{equation}
where $V$ is a product of nearest neighbor Givens rotations. This represents the correlation matrix in the basis $\tilde{\mathbf{f}}^\dagger = V \mathbf{f}^\dagger$. We identify the corresponding many-body operator $\hat{V}_i$ that implements a single rotation $\hat{V}_i \mathbf{f}^\dagger \hat{V}^\dagger_i = V_i \mathbf{f}^\dagger$ in terms of a 2-qubit unitary as 

\begin{equation}
    \hat{V}_i = \begin{bmatrix}
        1 &   & \\
          & V_i & \\
          &   & 1
    \end{bmatrix} \;.
\end{equation}
We show the circuit decomposition of a single $\hat{V}_i$ in Fig.\,\ref{fig:GivensDecomp}. The total unitary is obtained by multiplying the $\hat{V}_i$ in reverse order, \ie, if $V=V_1 \cdots V_m$ then $\hat{V} = \hat{V}_m \cdots \hat{V}_1$. Finally, we prepare the desired state $\ket{\Psi_F^I}=\prod_{i,n_i=1} \tilde{f}_i^{\dagger}\ket{0} $ by first initializing $\ket{\Psi_F^I}'=\prod_{i,n_i=1} f_i^{\dagger}\ket{0}$ via single-qubit $X$-gates acting on the appropriate sites, and then applying $\hat{V}$ (note that $\hat{V} \ket{0}=\ket{0}$:

\begin{equation}
    \hat{V} \ket{\Psi_F^I}' = \prod_{i,n_i=1} \hat{V} f_i^{\dagger} \hat{V}^\dagger \hat{V} \ket{0} = \prod_{i,n_i=1} V^{\hphantom{\dagger}}_{ij} f_j^{\dagger}\ket{0} = \prod_{i,n_i=1} \tilde{f}_i^{\dagger}\ket{0} = \ket{\Psi_F^I} \;.
\end{equation}
An illustration of the resulting circuit structure is shown in Fig.~\ref{fig:InitStateGivens}.


\section{Circuit compression}\label{App:compression}

\begin{figure}[]
    \centering
    \includegraphics[width=0.52\linewidth]{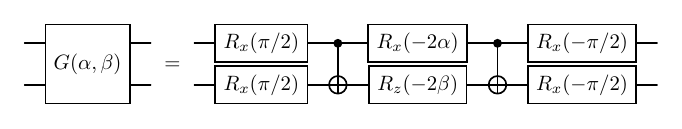}
    \caption{Decomposition of $G(\alpha,\beta)$ into single-qubit and CNOT gates.}
    \label{fig:Ggatedecomp}
\end{figure}

\begin{figure}[]
    \centering
    \includegraphics[width=0.4\linewidth]{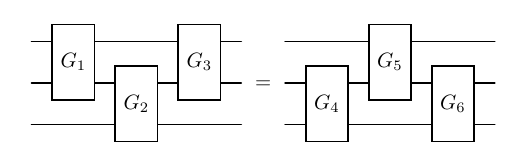}
    \caption{Turnover symmetry for two-qubit gates $G_i$ taken from a subgroup of U(4)}
    \label{fig:Turnover}
\end{figure}

\begin{figure}[]
    \centering
    \includegraphics[width=0.47\linewidth]{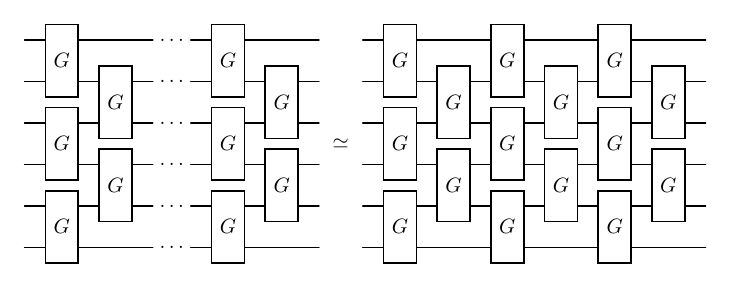}
    \caption{Compression of brick circuit of arbitrary depth to a fixed number of layers equal to the number of qubits by repeated application of the turnover symmetry.}
    \label{fig:brick}
\end{figure}

\begin{figure}[]
    \centering
    \includegraphics[width=1.0\linewidth]{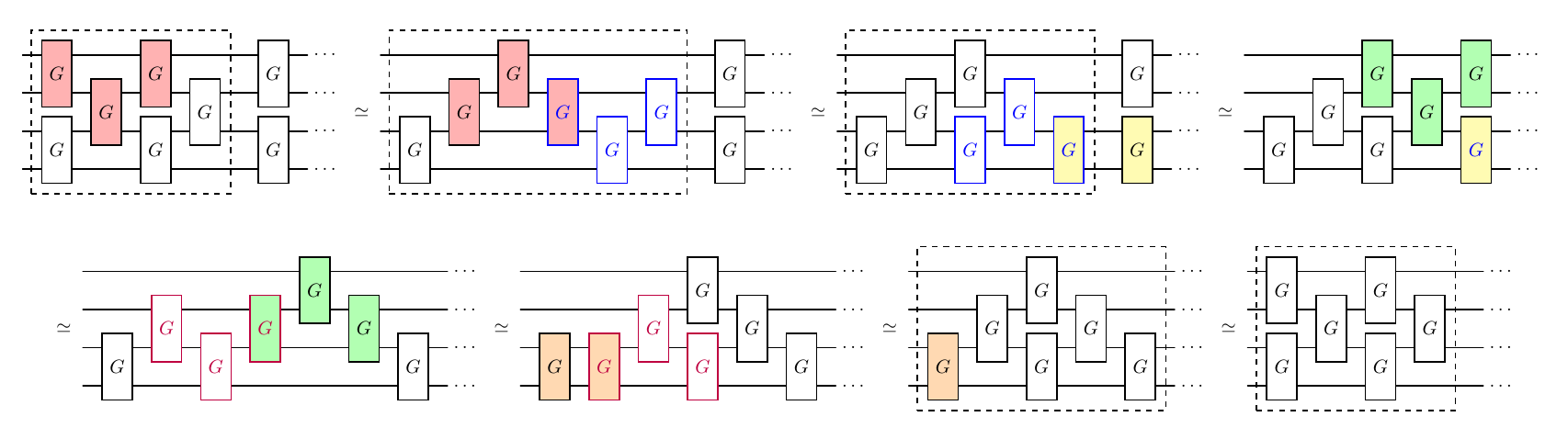}
    \caption{Sequence of application of the turnover identity that transforms an N-layered brick circuit on N qubits into a triangular circuit (dashed boxes) and then absorbs additional layers before transforming back into a brick circuit. The case N=4 is shown for illustration. Colors indicate groups of gates before and after the application of either a turnover identity or absorption into a single gate. The last step represents the reverse of the initial transformation from square to triangle circuit.}
    \label{fig:compression}
\end{figure}

We summarize the essential results underlying exact circuit compression in the context of the model studied in this work \cite{doi:10.1137/21M1439298,BassmanOftelie2022,PhysRevA.105.032420,f3c,qclab}. In order to implement the unitary time evolution governed by the local Hamiltonian $H_\mathrm{HCB}$ we discretize the exact unitary into time steps $\Delta t$,

\begin{equation}\label{eq:Trotter}
    U(n \Delta t) = \left( \mathrm{e}^{- \mathrm{i} H_\mathrm{HCB} \Delta t} \right)^n   .
\end{equation}
We then split the terms appearing in $H_\mathrm{HCB}$ into two sets of commuting operators, given by all $X_i X_{i+1} + Y_i Y_{i+1} $ with $i$ even and all $X_i X_{i+1} + Y_i Y_{i+1} $ with $i$ odd respectively. Operators from different sets generally do not commute, but as $\Delta t \to 0$ we have

\begin{equation}
    U(\Delta t) = 
    \left[ \prod_{\substack{\mathrm{i} \; \mathrm{odd} \\ \vphantom{>0} }} e^{i \Delta t \left(X_i X_{i+1} + Y_i Y_{i+1} \right)} \right]  
    \left[ \prod_{\substack{\mathrm{i} \; \mathrm{even} \\\vphantom{>0}}} e^{i \Delta t \left(X_i X_{i+1} + Y_i Y_{i+1} \right)} \right] + \mathcal{O}(\Delta t)   .
\end{equation}
This generalizes to the XY model $H_{\mathrm{XY}}=- \sum_i\left(J_x X_i X_{i+1} + J_y Y_i Y_{i+1} \right) $ with arbitrary real coefficients $J_x$, $J_y$. Each of the unitaries $g(\alpha,\beta) = e^{i \left(\alpha X_i X_{i+1} + \beta Y_i Y_{i+1} \right)}$ can be expressed as a two-qubit gate $G$ (for decomposition see Fig.~\ref{fig:Ggatedecomp}) and therefore \ref{eq:Trotter} corresponds to the brick circuit shown on the left hand side of Fig.~\ref{fig:brick}. Remarkably, there exists a symmetry for gates belonging to the group $G$ that is parameterized by $g(\alpha,\beta)$, namely the one shown in Fig.~\ref{fig:Turnover}, where the indices label different values of the angles $\alpha$, $\beta$. Given $G_1$, $G_2$, and $G_3$, one can always find $G_4$, $G_5$, and $G_6$ such that the two circuits are identical. Repeated application of this symmetry and the fact that two adjacent gates acting on the same pair of qubits combine into a single gate by adding the angles allows for the brick circuit shown on the left hand side of Fig.~\ref{fig:brick} to be compressed into a brick circuit of depth = \#qubits shown on the right hand side of Fig.~\ref{fig:brick}. We outline the sequence in which the turnover identity is applied in Fig.~\ref{fig:compression} for the case of four qubits as an example.

\end{document}